\documentstyle{l-aa}
\input{psfig.sty}
\begin{document}
\thesaurus{01
          (11.05.2: Cl0024+1654, A370;
           11.17.3;
           12.03.2)}
\title{Distribution of galaxies at large redshift and 
cosmological parameters from the magnification bias in  
Cl0024+1654\thanks{Based on
data obtained with the CFHT in Hawa\"{i}}}
\label{}
\author{ B. Fort\inst{1} 
        Y. Mellier\inst{2,1,3}
        M. Dantel-Fort\inst{1}}
\offprints{Y. Mellier
           mellier@iap.fr}
\institute{Observatoire de Paris, DEMIRM. 61 Av. de l'Observatoire, 75014 Paris, France
\and
        Institut d'Astrophysique, 98 bis Bd. Arago, 75014 Paris, France
\and
	Observatoire Midi-Pyr\'en\'ees, UMR 5572. 14 Av. Edouard Belin, 31400 Toulouse,
        France}
\date{Received May 1996, }
\maketitle
\markboth{Redshift  of sources and  cosmological parameters}{ from magnification bias in Cl0024+1654}

\begin{abstract}
We analyse the surface density of very faint galaxies
at the limit of the sky background noise 
in the field of the cluster  of galaxies  Cl0024+1654.
The radial variation of their number density
in the magnitude bins $B=26-28$ and $I=24-26.5$
displays an (anti)bias magnification effect similar to 
the one observed in A1689 by Broadhurst
(1995) for $I < 24$. The study of this effect provides
a determination of the radial distribution of critical lines
of the gravitational lens from which we infer the 
 redshift range of the populations seen in $B$ and $I$.

The surface density of $B$ galaxies drops rapidly to a  well-defined
inner critical line  near the large triple arc. The depletion 
extends from $R_B=30"$ to $R_B=53"$ and the shape of the curve 
can almost be reproduced with two redshift populations selected by 
the two filters. 
 With our data $60\% \pm 10\%$ of the 
 $B$-selected galaxies 
 are between $z=0.9$ and $z=1.1$ while most of the remaining $40\%$
galaxies should be at a redshift  close to
$z=3$.

The $I$ selected population observed with the lens 
has a similar but broader depletion 
with a minimum extending from the $B$  inner critical line to $R_I=60"$.
Whatever the cosmological model, the very faint $I$-selected galaxies  
spread up to a larger redshift 
with about 20\% above  $z > 4$. 
The fact that many faint $I$ selected galaxies
are not detected in B also favour the existence of a more distant
population with a redshift 
range between $z=3$ and $z=4.5$.

Using a model for the gravitational potential derived from a study of the 
large triple arc 
seen around the cluster, the locations of the two extreme critical lines for 
the B and I selected 
galaxies favour $\Omega_{\Lambda}$-dominated flat universes with
 a cosmological constant ranging from 0.6 to 0.9. 
The result is confirmed by a preliminary investigation of
the Broadhurst's effect in A370.  

 However, ultra-deep detection of faint distant galaxies
down to the  noise level
are technically very difficult.
In this first paper we mainly discuss the method to search the so-called "last
critical line" and several possible effects which may bias the 
results on cosmological parameters. 
We conclude that the systematic measurement of this outer critical line
of the faint $I$ selected galaxies population around
many clusters with gravitational arcs of known (low) redshift
may help to count the number of faint galaxies at very large redshifts 
beyond the possibility of any spectroscopic survey, and eventually to 
settle the issue of the existence of the cosmological constant. 

\end{abstract}

\section{Introduction}
Broadhurst et al. (1995) and Broadhurst (1995) 
 pointed out the interest of measuring the apparent variation of the 
surface density of background galaxies around massive clusters of galaxies. 
This variation is due to competing gravitational magnification and deviation 
effects.
It depends on the slope of the galaxy counts as a function of magnitude and 
on the magnification
factor of the lens which is responsible for
a radial stretching of the field of view around the
cluster.  The effect can be used  to map
the region where the magnification factor is
maximum (critical lines) and to evaluate the mass of the cluster.
More generally, it provides a way to break the intrinsic degeneracy
of the inversion methods in studies of gravitational weak shears
(Schneider \& Seitz 1995). Combined together, these tools 
provide a unique way to recover the total mass and the mass profile of 
gravitational systems.\par 
The radial
surface density of background galaxies up to a magnitude limit  $m$
can be expressed as
\begin{equation}
N(<m,r) = N_0(<m) \ \mu(r)^{2.5\gamma-1} \ ,
\end{equation}
where  $\gamma$  is the intrinsic count slope
\begin{equation}
\gamma = {dlogN(<m) \over dm }\ ,
\end{equation}
$\mu(r)$ is the magnification factor of the lens and $N_0(<m)$ the
intrinsic counts in the absence of a lens, which is obtained from counts
in a nearby empty field.
It is clear from expression (1) that 
a radial magnification bias $N(<m,r)$ shows up only when
the slope $\gamma$ is different
from the value 0.4; otherwise, the increasing number of
magnified sources is exactly cancelled by the apparent field dilatation
and there is no effect on $N(<m,r)$.
A radial amplification bias should not be observed behind a 
nearby cluster in $B(<26)$ since
the slope is almost this critical  value (Tyson 1988).
It is only detected in the R or I bands for which the slopes are close to 0.3
(Smail et al. 1995).

When $\gamma < 0.3$ a decrease of the number of galaxies
is expected in regions of magnification.
The effect is strongest at the critical radius around the cluster
corresponding to the most probable redshift of the 
background sources.
Since the critical radius increases with redshift, two populations
of different redshift will show two different radial depletion curves, with 
their minima deferring by an amount which depends on the redshift difference and
the cosmological model. Therefore, from the study of the radial
depletion, one can get information on the redshifts 
of the background galaxies. Alternatively,  Broadhurst
 (1995)  emphasised that this effect may be used to constrain the geometry
 of the Universe, provided we know the redshift of the two populations.

A difficulty comes from the fact that the critical radius 
writes $r_{cr} \propto D_{LS}(z_l,z_s,\Omega_m,\Omega_{\Lambda}) /
 D_{OS}(z_l,z_s,\Omega_m,\Omega_{\Lambda})$ and depends both from the redshift
 and cosmology ($D_{LS}$ and $D_{OL}$ are respectively the angular
distances from the lens to the sources and from the observer to the
sources). 
In this paper we show that it is possible to disantangle the 
redshift distribution from the geometry of the universe
provided the properties of the lens are known from
the  redshift of one critical line, such as the
redshift of a giant arc. In the case of Cl0024+1654
we show that  a population of faint blue galaxies exists that traces 
a reference critical line at low redshift.

There are only a few clusters for which the properties of the lens are known,
around which deep B and I observations are available on a large field.
An analysis of a  set of rich
lensing clusters is in progress  (Mellier, Fort \& Dantel-Fort 1996, 
hereafter paper II). 
In this paper, we present the results on Cl0024+1654 and some 
preliminary 
results on A370 in order to focus on the basic principles of the method.

In section 1 we show  that we can actually detect
in deep CFHT images very faint galaxies at the limit of the
noise in the respective bins $B=26-28$ and $I=24-26.5$ and
that the slope of count rates in both magnitude range are appropriate to 
produce 
an observable magnification bias.  This approach has been motivated by the 
recent
works done by van Waerbeke et al. (1996a) who used the pixel-to-pixel 
autocorelation function (ACF) of the sky background for measuring the shear
field around lensing clusters. 
Though the ACF does not need object
detection and cannot provide simply informations on the galaxy number
counts, from the strong signal they measured in
Cl0024+1654, we can predict that the number density of faint
sources is high and could be useful for the magnification bias.  

The second section discusses the depletion
around Cl0024+1654 (z = 0.39). This cluster is ideal because
giant multiple arcs have been observed which provide
consistent lens modelling 
 (Kassiola et al. 1992, hereafter KKF; Bonnet et al. 1994, 
Wallington et al. 1995). 

The last section discusses the results including some results obtained on
A370. 

\section{Deep galaxy counts  at $B > 26$ and $I > 24$}

Deep observations of the cluster 0024+1654 (z = 0.39)
and of a nearby reference field offset by about one degre from the 
cluster center were secured at the prime focus of the CFHT
 during the two nights in $B$ and $I$ with
a constant seeing  (0.7" in $B$, 0.55" in $I$). The
total exposure time are  10800 s for each filter.
In $B$ the field of view is 4.15'$\times$3.1' for the cluster fields and only 
2.2'$\times$ 3.5'
for the blank field. In $I$, the field of view is  7'$\times$7' for
the cluster and the blank field. The $I$-blank field observation
was contaminated by a strange meteoritic shower
(Jenniskens et al. 1993) and the corresponding counts is not considered.
The I count is instead extracted from 
the Hubble Deep Field study.
A description of the observations, with a modelling of the
large triple arc around the cluster, can be found in  
KKF. The ellipticity 
of the potential is small and we assume that it 
does not change appreciably within 
30" to 60" from the cluster center.
\par
We detected very faint galaxies
by using the SExtractor software (Bertin and Arnouts 1995).
The optimization of the input parameters necessary to  detect extremely 
faint objects  
was performed from  simulations and are available on request with 
 a reference blank field in $B$. The simulated galaxy counts
parameters of the observations were chosen
to be similar to those in our CFHT CCD images.
\begin{figure}
\psfig{figure={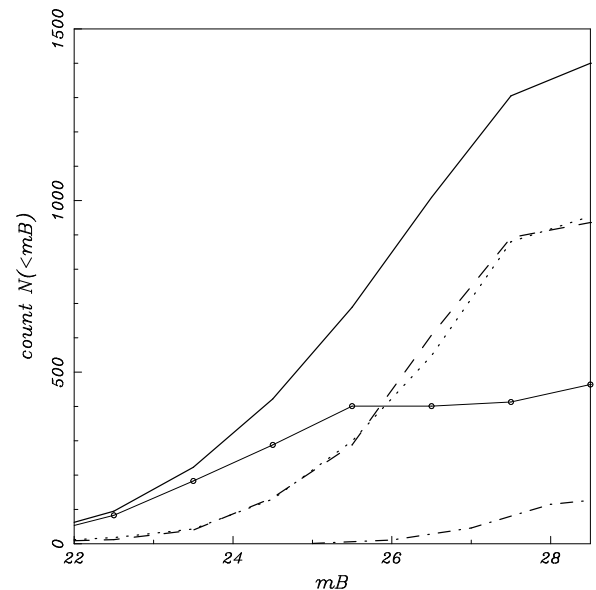},width=8.8cm}
\caption{Cumulative $B$ galaxy counts in the fields Cl0024+1654 (full
line), in the
CFHT blank field (dashed line) and obtained from the HDF (dotted line). 
The agreement between the CFHT blank field and the HDF is excellent in 
all the magnitude bins. The thin line with open circles show the cluster
members as expected from the subtraction of the cluster field to the
blank field. The flat shape beyond $B=25.5$ show that the contamination
is negligible in the magnitude range $26<B<28$ discussed throughout
this paper. The dotted-dashed curves show an estimate of the fraction of
spurious detections inferred from the optimized parameters used in
SExtractor on simulated CCD images. 
 }
\end{figure}

One of the main concerns is the significance of detection at
faint magnitudes. Since a wrong estimate of the count
slopes could change the estimate of the magnification, we
performed additional simulations
of a blank field, free from any objects (stars or galaxies), with only
a pure poissonian sky background.  We then optimised the parameters
defining galaxies 
so that only a small fraction ($<10$\%) of spurious events are
detected. Indeed the detection depends on the noise amplitude
and the method shall be used on images with good
flat-fielding and stable noise statistic. 
Any area with an enhancement of the background  like 
scattering of light around bright stars  or  bright galaxies 
will change the
limit of detection locally and shall be masked on the image.
 Finally, the surface number density inferred from the galaxy counts 
in Cl0024+1654  
fields has been corrected from the effect of sky occultation by
bright cluster members. 

Figure 1 shows the $B$  counts in the
cluster field and the blank field. The counts have been done
independently in the $I$ filter also. The  detection limits 
are respectively $28$ in $B$, and $26.5$ in $I$.  The number of expected 
spurious detections coming from simulations is also indicated.  
A plot of the recent Hubble Deep Field  counts (Couch 96, private 
communication) is also given. The agreement between the HST counts and
the CFHT counts for the blank field is excellent, giving confidence in 
our method.  This is a good exemple of what can be done from the 
ground with good seeing.

We find slopes of $0.17 \pm 0.02$ and $0.25 \pm 0.03$ respectively 
for the counts in $B$ at $26 <B<27.5$ and in $I$ at $24<I<26.5$ 
in the blank fields.
These values are close from the HDF counts (Table 1) and
the Tyson's counts (1988)
in $B$, and compatible with the deep I band counts of Smail et al
(1995).  One can see  on Figure  1
that at  $B > 26$ the contamination by faint cluster members
is negligible. 

\begin{figure}
\psfig{figure={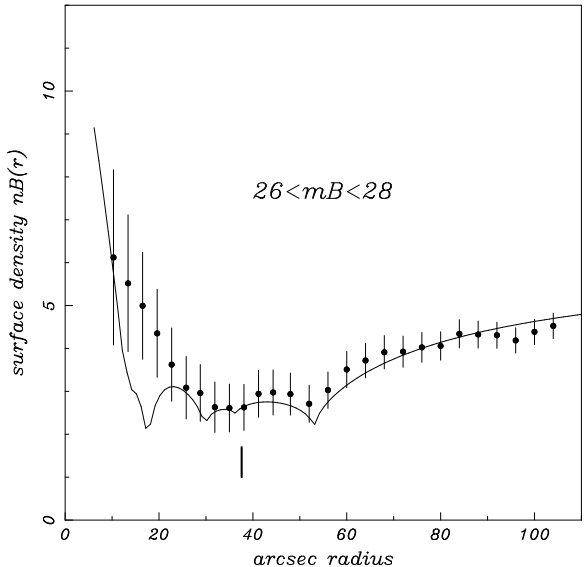},width=8.8cm}
\caption{Depletion curve obtained in Cl0024+1654 from galaxy counts in
the range $26<B<28$ for the $B$-selelected galaxies.
 The data are the filled circles with error bars. We
see that a depletion is detected, with a minimum starting
at
$R_B=30$ arcsec. The full line show a fit 
with the model
and core radius obtained by 
KKF. 
}
\end{figure}

\section{The magnification bias in Cl0024+1654 and the redshifts of the
blue and red background sources}

Since the galaxy counts in both $B$ and $I$ have slopes below  
0.25 we can expect a significant depletion of
galaxies at some radii around the cluster field. 
The radial surface density obtained for $B$ and $I$ are given on figure 2
 and 3 respectively. Both curves 
decrease with radius down to a minimum, and increase towards 
 $N_0$ at large radius. The $B$ shows a steeper descent. Its 
minimum is pronounced
at about 30 arcsec. and extends to 50 arcsec. from the cluster center.  
The $I$ curve shows a shallower broad dip ranging from 30" to about 60".  As 
the
$B$ and $I$ count slopes are almost identical, the width of the 
depletion probe the relative redshift distribution  of 
the two samples. The difference observed beyond 50 arcsec. already 
shows qualitatively that 
a  fraction of the $I$
 galaxies must be at a redshift significantly higher than the $B$ 
sample. 
Incidentlly we note that the  behavior of the
$B$-selected galaxies not detected in $I$ is similar to the whole sample
of $B$ galaxies.  

The depletions  of the curves given in figures 2 and 3 
can be reproduced from a  least square fit of  
 the depletion  generated by a population of galaxies located 
 at various redshift:
\begin{equation} 
N(<m,r) = \sum_{i=1}^{M} \ N_{0i}(<m) \ \mu_i(r)^{2.5\gamma_i-1} \ .
\end{equation}
where $ N_{0i}$, $\mu_i$ and $\gamma_i$  are respectively 
the number density of galaxies, the magnification   
and the slope of the galaxy counts of the population located 
at redshift $z_i$. The free parameters are
the slope of the galaxy counts, which at the starting point of
the minimisation are assumed to be  
the slopes of the blank field, the redshift and the number  density
 of the sources. 
 Though we allow the redshift distribution of galaxies
to be described with a maximum $M$ of 5 redshift  bins (equivalent to 5 
critical lines),  we searched for solutions using a minimum of lens 
screen to describe 
the distribution of sources in redshift which are selected through 
 our two filters. The coefficients $N_{0i}$ in
Table 1 correspond to the  galaxy number density at redshift $z_i$.
 It is important to note that
since the apparent depletion of galaxies only lies in  the innermost 
region of the lensing
cluster, the choice of the analytical function representing the mass profile 
is not critical for radius in the range 25 to 150 arcseconds 
(Bonnet et al. 1994).    
Therefore, we fit
the data by using the isothermal model of KKF with
a core radius $r_c=15$ arcsec and a velocity dispersion of 
$\sigma=1300 \ km/sec.$. 

In the following the position of the two extreme critical lines
are the important output of the fit rather than the relative redshift 
abundances of distant galaxies, a topic 
that would need more additional data on many clusters. Hence we have 
mostly explored
the possible range of variation for the radius of
the extreme critical lines in order to  estimate a possible
dependance on the cosmology.

\begin{table*}[t]
\begin{tabular}{lccccccc}
\hline
\multicolumn{8}{c}{Cl0024+1654 $B$} \\
\hline
$26<B<28$ & $N_{0tot}$ & $N_{0B}$ &  $N_{0bk}$  & $R_B$ & $\gamma_{HDF}$
& $\gamma_{fit}$ & z \\
\hline
All $B$ & 40 & 6.13 & 0.0 & --  & 0.17 & --  & --\\
Bin $B_1$ & -- & 2.60 & 0.0 & $30 \pm 3$ & -- & 0.20  & 0.90$^{+0.1}_{-0.1}$ \\
All $B_2$ & -- & 0.93 & 0.0 & $36 \pm 3$ & -- & 0.20  & 0.95$^{+0.1}_{-0.1}$ \\
All $B_3$ & -- & 2.60 & 0.0 & $53 \pm 3$ & -- & 0.20  & 3.00$^{+1.8}_{-0.5}$ \\
\hline
\hline
\multicolumn{8}{c}{Cl0024+1654 $I$} \\
\hline
$25<I<26.5$ &$N_{0tot}$  &$N_{0iI}$  & $N_{0bk}$ & $R_I$ &
$\gamma_{HDF}$ & $\gamma_{fit}$  & $z_l$ \\
\hline
All $I$ & 13 & 3.32 & 0.0 &  -- & 0.25 & --  &  -- \\
Bin $R_1$ & -- & 0.0 & 0.5 & $30 \pm 2$ & -- & 0.15 &
$0.90^{+0.1}_{-0.1}$ \\
Bin $R_2$ & -- & 2.1 & 0.5 & $32 \pm 2$ & -- & 0.20 &
$1.10^{+0.1}_{-0.1}$\\
Bin $R_3$ & -- & 0.06 & 0.5 & $41 \pm 2$ & -- & 0.24 & $1.30^{+0.2}_{-0.1}$
\\
Bin $R_4$ & -- & 0.53 & 0.5 & $48 \pm 2$ & -- & 0.20 & $2.30^{+1.0}_{-0.5}$
\\
Bin $R_5$ & -- & 0.63 & 0.5 & $60 \pm 2$ & -- & 0.2 &
$4.00^{+3.0}_{-1.0}$ \\
\end{tabular}
\caption{Summary of the galaxy counts and the magnification bias 
observed in the lensing cluster Cl0024+1654 in the $B$ and $I$ bands. 
$N_{0tot}$ is the 
total galaxy number density in the cluster field. $N_{0iB}$, $N_{0iI}$ 
are the true background components in the two bands, and the 
foreground or cluster members superimposed to the lensed population. The
densities are given in $gal/arcmin^2$ and are corrected from the
obscurations by bright foreground galaxies. The two last columns give
the slopes of the galaxy counts as inferred from the HDF and from the best
fit of the magnification bias. 
The   number density inferred when
the depletion is reproduced from a best fit is given for the various    
redshift bins found for the distribution of galaxies.  
The fit give $R_i$, $N_{0iI}$, and the 
slope $\gamma_{i}$. The redshifts of each bin does not depend much
on the cosmology and according to the discussion in paragraph 4 are given  
 for 
a flat universe  $\Omega_m=0.35$ and $\Omega_{\Lambda}=0.65$. }
\end{table*}

For the $B$ galaxies, the best fit of the depletion curve
can be modelled by using only three (almost 2) redshift bins for the $B$
sources population,  each of them
corresponding to a slope
$\gamma = 0.20 \pm 0.02$ in agreement with the slope derived
from the blank field. 
The results of the fit is  given in table 1 (reduced $\chi^2 = 0.66$). 
The inner critical radius is found  at  $R_B=29.8" \pm
0.4"$. This is a robust value which is well constrained by the steep
shape of the depletion as the radius increases from the cluster center. 
If we allow the core radius to vary around the KKF value we find a very similar
but marginally
best fit ($\chi^2 = 0.63$)
with a core radius of $13"$
and $R_B=30.2"$. The uncertainties coming from a 
small residual error on the value of the core radius of the KKF model 
is negligible and in 
the following we keep the KKF value.  But it is important to note
that both the core radius and ellipticity of such an isothermal model 
could be determined from
a mapping of the anti-bias magnification effect ($N(x,y; <m)$) over the whole 
 CCD image (see paper II).

Whatever the cosmology the KKF model imposes that the nearest 
$B$ galaxies have a redshift between $0.9$ and
$1.1$.   
In fact this value was not unexpected.
First, Mellier et al. (1991) obtained spectra of the 
images of the arcs whose postions are close to
$R_B=35"$ (see figure 2). They are blue objects and they should belong to 
the $B$ population.  The presence of
bright starburst spots clearly visible in the HST images of these arcs
(Colley et al. 1995) is a strong argument that star formation is ongoing in 
this galaxy.  Although a definite redshift cannot be given, 
the spectra provide compelling evidence that it could be close
to one, in any case larger than 0.88: otherwise we should see a proheminent
[OII]$\lambda$3727 emission line in the observed spectral 
range covered by the data.  Second, with the shear profile in Cl0024+1654,
 Bonnet et al. (1994) infer that the $B$ galaxies
responsible for the shear pattern should have an averaged redshift 
between 0.9 and 1.2. Third, the redshift survey of giant arcs 
(see Fort and Mellier 1994, Soucail 1995) and arclets 
(Bezecourt and Soucail 1996, 
Pello et al. 1996), and the lensing inversion in A370 and A2218 as well
(Kneib et al 1994, 1996) provide  a fairly large sample of
blue-selected galaxies with 
$B$ magnitude ranging from 24.5 to 27, which all have a  
median  redshift of 0.85. Most of them are between 0.65 and 1.10, but 
the lower redshifts of arclets
 have been found behind the lower redshift clusters ($z_l<0.25$) as 
expected from the properties of gravitational lensing. Besides,
 the average magnitude of these samples is about one magnitude brighter 
(26-27.5 instead of 24.5-27) so we can reasonnably expect that our faint
 $B$ galaxies should be somewhat more distant than $z=0.75$, whatever the
evolution model.

The depletion curve of the $B$ galaxies raises again at
$R_3 = 53"$ corresponding to a redshift of about  $z=3.0 \pm 0.5$
 well determined from the best fit. The redshift error   
is mainly dominated  by the uncertainty on the redshift of the inner 
critical line. An important point is that the position at which the 
$B$ depletion curve starts to increase  does {\sl not} depend on the 
number of source screen (Table 1), but only depends on 
the largest redshift screen. Therefore, the depletion curves has the
remarkable property to provide the spread of the redshift distribution,
from the width of the depletion,   
{\sl and} the highest redshift of the sample from the  comparison of the 
radius of the upper critical line with a lower critical line.
 At this stage and before the detailed discussion of extreme redshift in
paragraph 4 let's assume
that the blue population range from $z_B=1. \pm 0.2$ to 
$z_B=3$, with a relatively well defined lower limit from the 
convergence of the lens.
\par
\begin{figure}
\psfig{figure={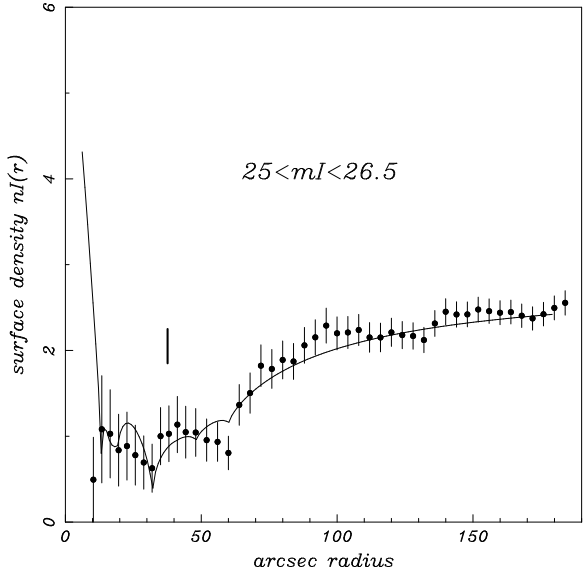},width=8.8cm}
\caption{ Same plot as figure 2 for the $I$ sample. 
The width of the depletion spread on a larger area.
The minimum radius is almost similar to the $B$
galaxies. The {\sl last critical radius} is defined as the position
where the depletion curve starts to increase. This corresponds to
$R_I=60"$ arcsec.
}
\end{figure}

For the  faint $I$ galaxies the observed  number density   
$N_I(25<I<26.5)$ is depleted in a larger range of radii.  The beginning
of the dip for the $I$ galaxies is almost the same as for the $B$
galaxies, thus their averaged redshift is the same (Table 1). This is not
surprising as this is the lowest redshift for which the gravitational
magnification is substantial. The minimum of the depletion is 
broader than for the $B$-selected sample but it has a bump between the two
extreme radii. 
In figure 3, the best fit of the
data uses the same lens model as for the $B$ galaxies, but needs a 
model distribution of galaxies with 4 discrete different $z$   
corresponding to  critical lines $R_i$. 
The reduced $\chi^2$ is very good (0.57)
if we substract to the data a uniformly distributed component 
corresponding to 14\% of the count in the blank field. The origin of this 
small count excess  and bump in the bottom of the depletion  
is not clearly understood so far (cf paragraph 5). 
If the bump and the relatively deep dip 
near the last critical line are real (Figure 3) they are not 
perfectly reproduced 
with the model (table 1).
Instead the radius of the two extreme critical radii  are robust
and the best fit  gives: $R_2 = 32 " \pm 1"$ and $R_5 = 60" \pm 2"$.
From the modelling of the flat depletion curve we 
conclude that  
the redshifts of the $I$ galaxies are $1.<z_I<4.5$, but with a significant
fraction at large redshift (table 1). 
According to the KKF model 
the most distant critical line corresponds to an abundance of galaxies
of $20\%$ at $z>4$.
This is an extreme critical line  where 
the effects of the cosmological parameters can be seen. In fact
Cl0024+1654 is an ideal configuration  where the role of the cosmology can 
be observed
directly from the radius of an extreme critical line, though its effect 
is mixted with the redshift distribution 
of the sources.  In the next section we 
consider in detail the possibility to obtain simultaneously the redshift
distributions and the cosmological parameters from the positions of
critical lines.

\section{A tentative to constrain the cosmological parameters with Cl0024+1654}

\subsection{The relative scaling method}
The beginning of the dip of the  $B$ depletion and the radial position at 
which the number density of $I$ galaxies  raises again define two extreme
 critical lines, $R_{min}$ and $R_{max}$. Their ratio writes:
\begin{equation}
{R_{min} \over R_{max}} = 
\left(
{
 \left( {D_{LS} \over D_{OS}} \right)_{max}^2 
 \left( {6\pi \sigma^2 \over c^2}  \right)^2  - r_c^2
\over 
 \left( {D_{LS} \over D_{OS}} \right)_{min}^2 
 \left( {6\pi \sigma^2 \over c^2}  \right)^2 - r_c^2
} 
\right)^{1/2}
\end{equation} 
and provides an interesting observable parameter based on
the two well-measured positions of the smaller and the 
larger critical radii. The dependence on  
$\sigma$ is  weak and is non-existent  for a singular isothermal
sphere. The dependence with $r_c$ is generally weak for simple potential
since the  cluster core radii are small with 
respect to the position of the external critical line (see Fort \& Mellier 
for a review).  However, it depends on 
$z_B$, $z_I$ and the geometry of the universe defined by $\Omega_{m}$
and $\Omega_{\Lambda}$. 
 
It is possible to map the volume space ($z_{min},z_{max},\Omega_{m},
\Omega_{\Lambda}$) in order to 
find the solutions which gives a $R_{max}/ R_{min}$ in agreement with the 
observations.  The results are
shown on figure 4 for flat universes, and on figure 5 for universes
with  a cosmological constant equal to zero.
 We have kept all the acceptable solutions, with extreme error 
bars ($R_{Bmin}= 30" \pm 0.5"$, $R_{Imax}=60" \pm 2"$), which corresponds to 
$1.9<R_{max}/R_{min}<2.1$, for 
an isothermal models with a core radius corresponding to the best 
fit parameters found by KKF. The
plots only show models with $z_B=0.85, \ 0.95$ and $ 1.1$, because they
are sufficient for the conclusions.  Both 
figures show that we can find acceptable solutions for any cosmological
models. However, there are clear trends for the redshifts of the sources.
Whatever the acceptable models, they all
imply that the clostest $I$ and the $B$ galaxies are at redshift {\sl lower} 
  than 1.0, and
the most distant $I$ galaxies are at redshift {\sl higher} than 2.5. 
 A  limit of the $I$ galaxies ($z_I<6.5$) 
is given by 
the filter itself which imposes that these galaxies have the
$Ly_{\alpha}$ break below 9000\AA. A lower limit at $z=0.85$ is 
given by the spectroscopic informations from the giant arc. 

The relative scaling method gives informations   
on the redshift distribution of galaxies that are almost independant from
the lens modelling. 
But it does not constrain  much the
 cosmology as far as the information on the exact
radial location of the critical lines is  not used. 
These positions  depends  on the lens 
model through the parameter $\sigma$. If $\sigma$ is determined
from the redshifts of arcs located at low redshifts,  
the location of the most distant $I$ critical line becomes 
a relevant observation for  the cosmology.
\begin{figure}
\psfig{figure={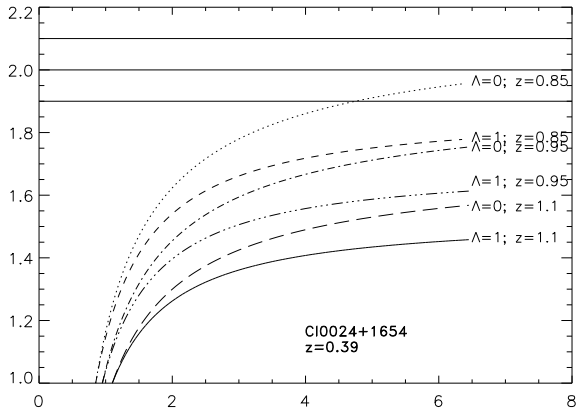},width=8.8cm}
\caption{Ratio $R_I$/$R_B$ versus redshift of the $I$ sources for three
redshifts of the blue populations ($z_B$) corresponding to the 
beginning of the dip in the $B$ depletion curve, and various flat universes. 
The horizontal lines give the same quantity as inferred from the
observations of $R_B=30 \pm 0.5$ and $R_I=60 \pm 2$ and computed with a 
singular isothermal sphere with a core radius of 15 arcsec and a
velocity dispersion of 1300 km/sec. The central line is the central
value and the two other lines show the position at $\pm 1 \sigma$. Since
we only use ratios of angular distances, it does not strongly depends on
the lens modelling. 
We see that the data are compatible with any models, but
the redshift of the blue population must be below $z_B=0.95$.
 }
\end{figure}
\begin{figure}
\psfig{figure={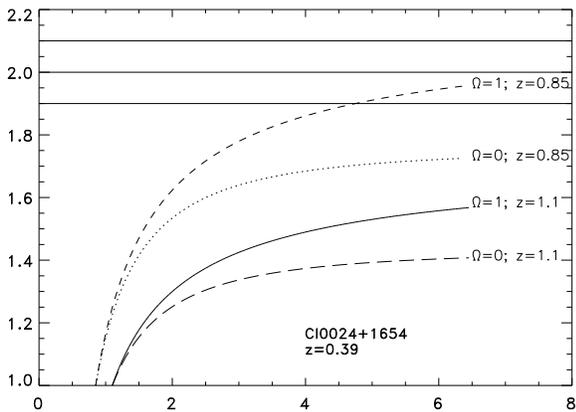},width=8.8cm}
\caption{Same plot as figure 4 but for cosmological models with a
cosmological constant equal to zero. Again any value of $\Omega_m$ is
compatible with the observation, provided the redshift of the blue
population observed at the beginning of the dip is lower than 0.95.
 }
\end{figure}

\subsection{Scaling the critical line with the lens modelling}
 From the last critical line $R_I$ which lies at a large distance
from the cluster center
it is possible to compute the angular distance ratio $D_{LS}/D_{OS}$ 
with the lens model. For simplicity, let us just
express it for the case of an isothermal
sphere with a core radius $r_c$. The relation between  $D_{LS}/D_{OS}$ 
 and the position of the  critical line $R_I$ is:
\begin{equation}
\left( {D_{LS} \over D_{OS}}\right)_I
 =  \left(R_I^2+r_c^2\right)^{1/2} \left({c^2 \over 6
\pi \sigma^2} \right) \ .
\end{equation} 
For a given $\sigma$, this ratio becomes significantly dependent
 on the cosmological model when the
source redshift is large, as it is the case for the $I$ population. 
Hence we search in which redshift range and for 
which set of  cosmological parameters, the 
ratio $D_{LS}(z_I,\Omega_m,\Omega_{\Lambda})
/D_{OS}(z_I,\Omega_m,\Omega_{\Lambda})$ is equal to 
the observed value.

\begin{figure}
\psfig{figure={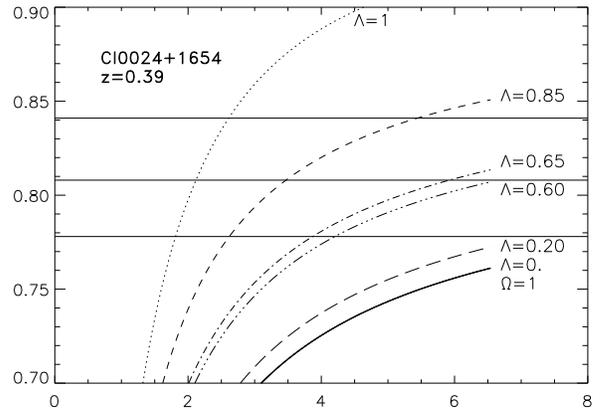},width=8.8cm}
\caption{$D_{LS}/D_{OS}$ versus redshift for the I population in
Cl0024+1654 and various cosmological models. The curves are strongly
dependant upon a scaling factor which should be given by the accurate redshift 
of a large multiple arc. Here the horizontal lines give
the value infererred from the observation by using an isothermal sphere
with core radius and the parameters of the best KKF model (full  central
line) corresponding
to a change of 0.1 in the redshift population of the blue galaxies (top
and bottom line). The ($\Omega_{\Lambda}=0,\Omega=0$) curve is the upper limit
for any models with null cosmological constant. Therefore, they seems all
ruled out by the observations. Low-$\Omega_{\Lambda}$ models are
marginally compatible with the observations 
but would require very high redshift for the $I$
galaxies. Best models are within the range $0.6<\Omega_{\Lambda}<0.85$.
 }
\end{figure}
Figure 8 shows  the relation $D_{LS}/D_{OS}$ to the redshift of the
sources, for a lens at $z_l=0.39$. The curves correspond to various 
cosmological models with a zero cosmological constant, or with a zero  
curvature. The horizontal line gives the observed ratio for the 
last critical line of the $I$ population in Cl0024+1654 ($R_I=60"$), 
computed with the best parameters of KKF ($\sigma=1300 km/sec$). 
We see immediately that many models are rejected. In particular there 
is no model with a null cosmological constant compatible with the data.
This seems to rule out any open universe and a ($\Omega_m=1$)-universe 
as well. On the
other hand, models with cosmological constant {\sl larger} than 0.65 are 
acceptable. The upper limit for the cosmological constant is not
provided by the lens, but some conservative upper bounds are given from
the $\Omega_m$ inferred by the CNOC dynamical analysis of rich clusters
(Carlberg 1996) which give $\Omega_m =0.16 \pm 0.2$. For a flat universe, 
this gives $0.65 < \Omega_{\Lambda} < 0.85$ . Furthermore, if we consider 
the robust  upper limit of Kochanek (1996) coming from
 the statistic of gravitationally lensed QSOs, 
the possible range is apparently converging on an almost unique 
value:  $ \Omega_{\Lambda} \approx 0.65$. 

The redshift range permitted for the distant $I$ population is
between 2.5 and 6.5, in agreement with the previous relative scaling method. 
The redshift of the population at the first critical line observed at
the beginning of the dip is still found  to be near $0.9
\pm 0.1$ as expected from the discussion in  3 and 4.1.
\par
In this method, the determination of $R_I$ is the critical parameter.
 The  uncertainties are dominated by the lens model, particularly 
by the  error on the velocity dispersion. Assuming a redshift between 
$z=2$ to $z=1$ for the giant arc, 
the KKF's best model gives $\sigma$ between 
1200 and 1400 km/sec, whereas  Wallington et al. 
(1995)  give 1100 to 1400 km/sec. This uncertainty will be
 suppressed as soon as the redshift of the triple arc will 
be well known (an observational priority). Both determinations are  
in good agreement with the galaxy velocity dispersion directly measured by
Dressler et al. (1260 km/sec; 1985). In fact, it is remarkable that 
Bonnet et al. have 
already shown that the shear profile in Cl0024+1654 is compatible with a 
isothermal sphere with $\sigma=1300km/sec$ and $z_{sources}=0.9$. If 
the velocity dispersion we 
use from the lens modelling is correct, then the uncertainty only comes from 
the rather small redshift
range we found for the population at $R_{min}$ which can be converted to an 
equivalent small error on the
velocity dispersion. For $\delta z_{min} \approx 0.1$, $\delta \sigma$  
is as low 
as 50 km/sec. On figure 7 we have
also plotted the observed ratio $D_{LS}/D_{OS}$ for $\sigma=1350$ km/sec
and $\sigma=1250$ km/sec.  One can see that the constrains are still 
reasonnable, but it clearly emphasises that the dependance on $\sigma$
may be an important shortcoming. 

Waiting for  
a precise redshift measurement of the large arc in Cl0024+1654 arc 
one have tentatively try to check the consistency of the
results by searching the last critical line of the $I$ population on 
another cluster  for which the redshift of large multiple arc is known:
 A370.

\begin{figure}
\psfig{figure={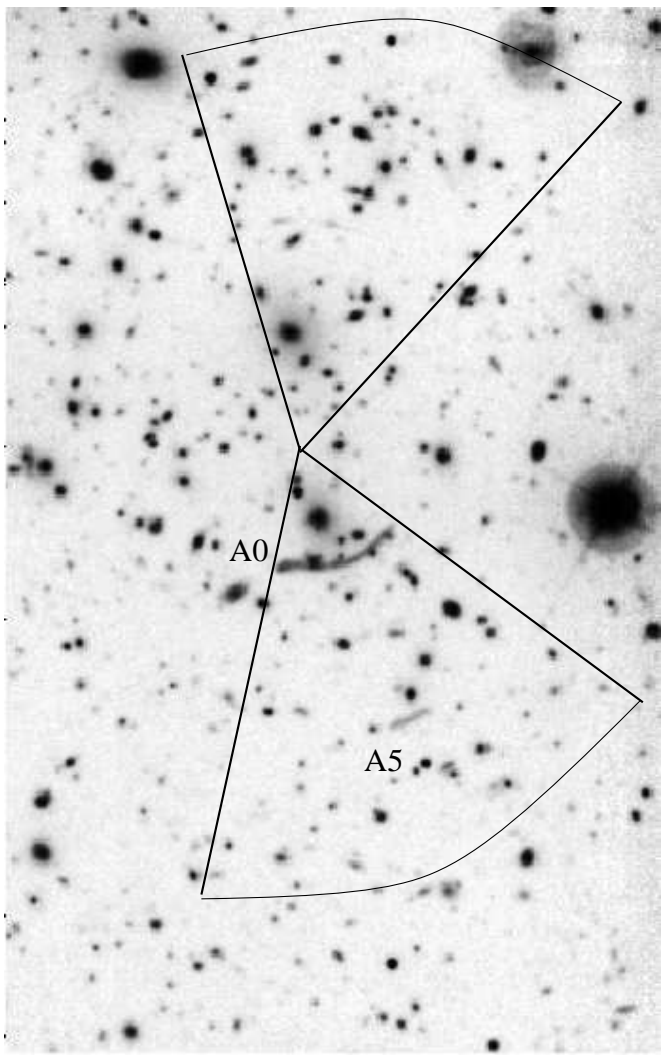},width=8.8cm}
\caption{CCD image of A370 showing the delimited areas where the
counts have been performed. These are regions where crowding by
bright galaxies is small and where no additional mass perturbation 
is detected. 
 }
\end{figure}

\subsection{Scaling the last critical line in A370}

The redshift of the arc is 0.725 (Soucail et al. 1988) and the galaxy
velocity dispersion is 1300 km/sec (Mellier et al. 1988). The cluster is
at a redshift of 0.375, almost identical to Cl0024+1654. 
Unfortunately,
A370 is not as regular as Cl0024+1654. Its bimodal structure is 
well established from the lens modelling (Kneib et al. 1993) and from the 
ROSAT X-ray map (Fort and Mellier 1994). Furthermore, the arclet
distribution shows that there is probably another mass condensation
to the east (Kneib et al 1994).  We restrict our study to 
two $60^o$-sectors along the axis passing through
the center of the two cD galaxies which appear almost free of additional 
perturbations and are 
not crowded by bright foreground galaxies. They are shown on figure 9. 
Since we only have  very
deep $I$ images with very good seeing, we concentrate on this color only. 
A detailed analysis will be given in paper II and we just give here
a preliminary analysis that uses the 
symmetry of the bi-modal potential. 

The two clumps have a remarkable alignement and almost similar 
geometrical and dynamical parameters (Kneib et a. 1993).
 The depletion curves can be refered to the barycenter of the two
clumps, though the bimodal shape of A370 makes the relation between the
last critical line and the redshifts of sources less obvious than
Cl0024+1654. 
Since we have selected two symmetric areas aligned with the main axis of 
A370, we expect
the critical lines to be almost the same in the opposite North and South 
sector cones.  The  depletion curves are plotted on figure 10.
 They are very noisy 
because the number of detected galaxies is much lower than around Cl0024+1654.
Indeed they do not display a broad minimum so clearly, though their shape 
 reveals a depleted area similar to  Cl0024+1654.
 We find the possible radius of the extreme critical line along
the main axis to be at $60 \pm
5$ arcsec, surprisingly similar to the value obtained for Cl0024+1654. 
The vertical line on figure 10 indicates the position of the giant arc 
critical line $R_B (z=0.725)$ along the main axis.

We applied the same method as in Cl0024+1654 to find the cosmologies
compatible with the data. 
The two cD potentials are identical, with the same core
radii $r_c$ and velocity dispersions $\sigma$. Hence, along the 
main axis, the potential well can 
be written:
\begin{eqnarray}
\Psi(x,y) & = & {D_{LS} \over D_{OS}} {6 \pi \sigma^2 r_c\over c^2}  
      \left(\sqrt{1+(1-\epsilon)(x-x_0)^2+y^2} \right.+  \nonumber  \\
 & &  \left. \sqrt{1+(1-\epsilon)(x+x_0)^2+y^2} \right)
 \end{eqnarray}
where $\epsilon$ is the ellipticity of the two clumps (identical), and
$(x_0,y_0=0)$ gives their position with respect to the barycenter. 

 From $\Psi(x,y)$ one can  derive the position of the tangential
critical lines along the main axis.
This gives a relation between  the velocity dispersion, core radius
and scaling coefficient $D_{LS} / D_{OS}$  
\begin{eqnarray}
{D_{LS} \over D_{OS}} {6 \pi \sigma^2 \over r_c c^2} 
\left( { 1 \over \sqrt{1+(1-\epsilon)(x-x_0)^2}} \right.+  \nonumber  \\
\left. {1 \over \sqrt{1+(1-\epsilon)(x+x_0)^2}}  
\right) &  = & 1 
\end{eqnarray}
 which can be solved numerically. 

The term $6 \pi \sigma^2 / r_c c^2$ can be obtained directly by 
using equation (7) at the position and the redshift of the giant arc. 
 With a simple model assuming that $\epsilon$ does not change with 
radius, equation (7) can
be plotted in order to find acceptable solutions for the last critical
radius observed in A370.

If it is true,  we found similar constraints  on the redshift of sources 
and on the cosmological parameters as in Cl0024+1654.  The most distant
$I$ galaxies corresponding to the last critical radius have $2.7<z<5$ 
whatever the cosmological model. 
 More precisely, if $\Omega_m$ is larger
than 0.16 from the CNOC works (resp. 0.3 from weak lensing in clusters) then 
the $I$ galaxies are at a redshift lower than 6 (resp. lower than 4.7). 

Models with zero cosmological constant and $\Omega_m<0.15$ are
incompatible with the data. However,  we cannot rule out
any  values  for $\Omega_{\Lambda}$, even those close to $1$. 
A similar remark has been incidently 
made  by Kneib et al. (1993) when trying to
explain the large ellipticity of the arclet A5 in A370 with the lens
modelling. The uncertainty probably comes from the general shape of the
cluster which is not as regular as our analytical description. Fort \&
Mellier (1994) show the isoluminosity contours of the red cluster
galaxies and the X-ray map. If light traces mass, then the discrepancy with 
respect to an regular elliptical potential is clear. In this case, it is
more appropriate to map the magnification from the {\sl local} depletion
 $N(x,y,<m)$ everywhere in the cluster field. This preliminary study of A370 
confirms  that the method of
the last critical line  gives reliable result only if it is used 
in conjonction with a self consistent modelling
of the large arcs (paper II).

\begin{figure}
\psfig{figure={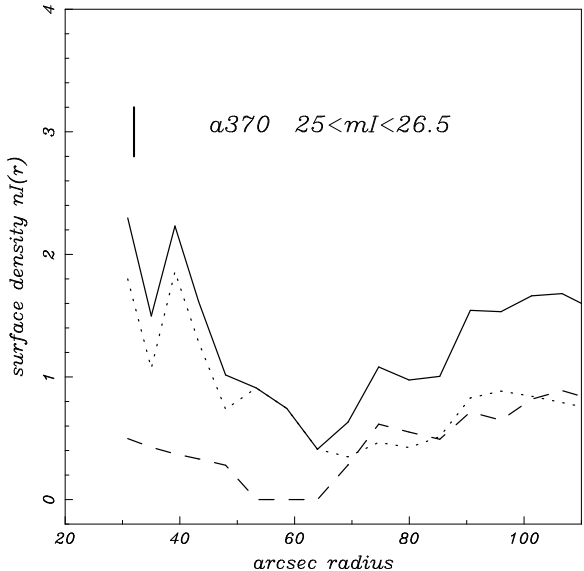},width=8.8cm}
\caption{Depletion curve similar as figure 2 and 3 obtained from the deep
I counts in A370. The dotted line show the count for the nothern region
only and the dashed line for the southern cone. The full line is the
addition of the two regions. The vertical line show the position of the
giant arc which provide and accurate image of the blue critical line. 
The depletion is clearly visible and shows the same characteristic as
the $I$ curve for Cl0014+1654, though the signal to noise ratio is
smaller. In particular, the last critical line is $60 \pm 3$ arcsec. 
 }
\end{figure}
\begin{figure}
\psfig{figure={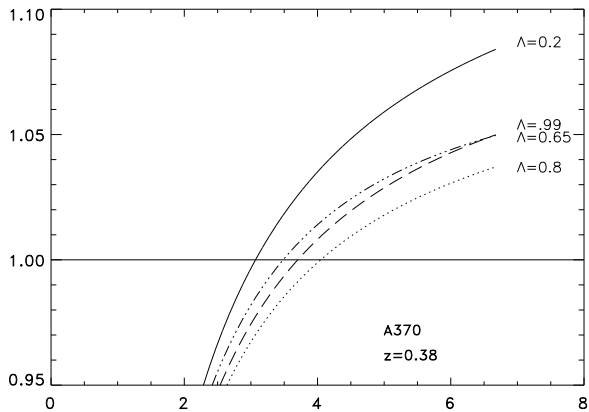},width=8.8cm}
\caption{Ratio $R_I/R_B$ versus redshift of the $I$ sources in A370 
for various flat universes with cosmological constant. The scaling is 
given  by the redshift of the giant arc ($0.724$), so we do not need to
plot curves for various redshifts as for Cl0024 (see fig. 5). The
horizontal line gives the same quantity as inferred from the observations
of $R_I=60$ and by using the simple model provided by equation (7).
There is no model ruled out by these observations, but they show again
that the redshifts of the $I$ population observed on the last critical
line should be larger in the range $3.-4.5$. 
 }
\end{figure}
\begin{figure}
\psfig{figure={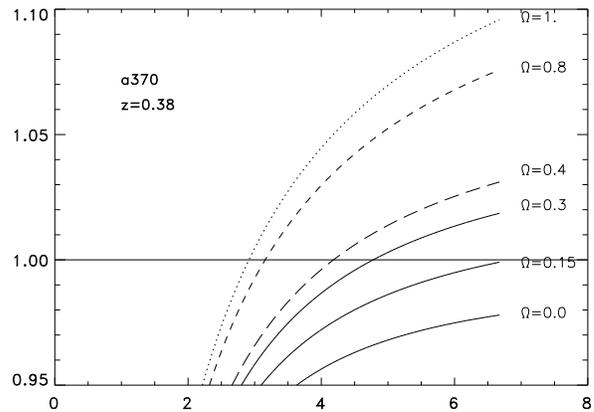},width=8.8cm}
\caption{Same plot as figure 9, for ($\Omega_{\Lambda}=0$)-universes.
The redshift of the $I$ population observed on the last critical line 
has a well-defined lower limit at $z=3.$, but the upper limit is not 
well constrained. However, redshift larger than $5$ are unlikely since 
$\Omega_m>0.15$ from the mass estimates of clusters from the CNOC
survey.  From the mass estimates obtained by strong and weak lensing
analyses in cluster, $\Omega_m>0.3$ and we can reasonably conclude that
$z<4.7$. 
 }
\end{figure}
\section{Discussion} 

The analysis of the magnification bias proposed 
by Broadhurst (1995) around rich lensing clusters 
 can be used to infer the redshift  distribution of  
faint galaxies. First tentatives in Cl0024 and A370 show that 
consistent informations were provided by these two clusters. 
The beginning of the dip at the first critical line corresponds to a 
redshift of $0.95 \pm 0.1$ for the $B$ and $I$ galaxies. 
 The magnified $I$ galaxies spread over a larger redshift range 
from $\approx 0.9$ to probably $>4$ in both Cl0024 and A370. 
From the study of Cl0024+1654 
we suggest that about $20\%$ of the faint $I$ population 
is at redshifts above $z = 4$,  and together with the knowledge of the lens 
model, namely the velocity dispersion, they give interesting information
 on the cosmology.
 It is surprising to find that both for the $B$ and $I$ selected galaxies 
 the abondance of galaxies is not a uniformly decreasing function of redshift
 but rather corresponds to
 a bimodal distribution with  few galaxies in the redshift bin
 $1.3-2$.  The shape of the distribution is remarkably similar to the 
photometric redshift 
distribution of the HDF population predicted by Stephen et al. 1996.  Our 
sample 
cannot reproduce the shape below $z=0.4$ for obvious reasons, but we 
clearly have a peak below $z=1.1$ and a secondary broad peak between 
$z=1.5$ to $z>3$.
 The apparent bump in the middle of the depletion curve in $I$ 
could be real bias magnification signature  but it could also be related to the
presence of an additionnal condensation of galaxies.

The use of the velocity dispersion obtained from the best lens modelling
of Cl0024+1654 gives the position of the inner critical
lines with a redshift consistent with our findings
from the spectro-photometric observation of blue arc(let)s. 
The spectrum of the Cl0024+1654 arc  
is also a preliminary
confirmation of the redshift prediction. For the most outer critical
line the  significant 
fraction of the red galaxies which are  {\sl not} detected in the $B$-band 
suggests that the $Ly_{\alpha}$ break lies beyond 5000\AA,    
which also favors a redshift larger than 3. These galaxies
might be  the faint end of population detected 
with the Keck telescope  by Steidel et al. (1996). 
From table 1 there number density is about $4 \pm 3$ /arcmin$^2$.  

For the cosmological test with the last critical line, a difficulty when
using the magnification bias from galaxy number counts is the 
clustering of the background galaxies that may adulterated 
the curve 
$N(r)$. Highly clustered background sources may bias the depletion
curves toward a well-defined peak at some radius. It seems  
this major difficulty has been mostly  avoided by going to fainter
background galaxies. An additionnal test of the surface brightness 
to  select the most distant ones with less clustering (see section 3) 
is too difficult for objects close to the noise.
The fact that the two curves obtained here fit well those 
expected from the lens modelling seems an a-posteriori confirmation
that the method is efficient. Besides, 
a large condensation of blue galaxies just behind the cluster center
 is not visible as a  bump in  the galaxies count in 
the cluster field with respect to our blank field 
or to the HDF field. 
Figure 1 does not show any perturbation of the slope in the cluster 
field.

The constraints  we obtained for the cosmological parameters 
 are essentially based on 
the determination of the "last" critical lines in $I$. The possibility
to materialize a critical line at very large redshift seems to provide 
 {\it a 
lower limit on the cosmogical constant} since the deviation angle is larger
than expected for an open universe.
Some effects may affect the results. 
One is the fact that we are observing clusters with large arcs
 which could be sources  associated with 
 a large condensation of mass also. This mass might  act as  an additional 
deflector  for most distant $I$ galaxies  that would then be 
deflected twice when the light passes through the two consecutive lenses. 
This might be wrongly  modelled as a single lens with a larger critical
radius. With deeper exposure time with a VLT a test can be performed
to detect such possible  condensation of galaxies by using 
color index distribution of galaxies (Mellier et al. 1994).
The frequent discovery of unexpected shear around foreground groups
in the field of over-luminous quasars (Bonnet et al. 1993,
Fort et al. 1996) is an indication that mass condensations are not 
rare. Similarly the "elliptical
crisis" that come from  the too high occurence of four lensed 
quasars can be also explained if their field has 
an additional shear field. These additional deflectors are indeed
particular cases of additional mass sheets which increase the angular scale 
of the gravitational bench as a cosmological constant would do. Such 
intervening matter could be difficult to observe since it does not
modify the shear pattern (Schneider \& Seitz 1995).
 A second possible bias is that the ellipticity
 of the cluster potential changes significantly between 30' and 60" from 
 the cluster center
 but this is not seen on the shape of the depleted area around Cl0024+1654.
 Although these two biases cannot be systematic for all cluster with
 large arcs  they have to be kept in mind for Cl0024+1654 and A370 which
 display several bright blue arclets. With the lens model it is possible
 to infer their clustering properties in the source plan if we suppose
 they are all at the same redshift. This should be done in the future.

 In conclusion, going close to the noise to search
 for an extreme critical line can provide information on the redshift 
distributions of the faintest galaxies and a direct test on the cosmology. 
 It seems from this preliminary investigation that a flat universe
 with a cosmological constant near 0.65  could be a possibility. 
 This value is still 
in agreement with the Kochanek's (1995, 1996) upper value coming from the
statistical analyses of lensed quasars, and from the tentative predictions of
Mo and Fukugita (1996).   It is also compatible with the parameter range
inferred by van Waerbeke et al. (1996b) from the neo-$V/V_{max}$
statistics of quasars.
As well, the result for a large abundance of
 high-redshift galaxies is compatible with the recent spectroscopic  detection 
with the Keck telescope by Steidel et al (1996). 
An interesting consequence of the large radial distance of the last
 $I$ critical line is that 
red gravitational pairs candidates should be now searched for 
  much larger radii than the large blue arcs.
  A few 
candidates are known around  some clusters and their spectroscopy could 
give a confirmation of all the results presented here. 
It is also clear that it is 
absolutely necessary to use our  method on a large numbers of clusters 
and to use very good modelling
of large gravitational arcs systems before to
expect strong constraints on the cosmological parameters. Such 
an observing programme is a typical one for 10-meters telescopes.

\acknowledgements{We thank James Lequeux, G. Mathez and Peter Schneider 
for useful
discussion, their encouraging support and comments on the 
manuscript. W. Couch kindly proposed his 
HDF catalog and his own selection parameters for running SExtractor. We
thank R. Pell\'o for providing the redshifts of arclets in A2218 prior to 
publication.}

\end{document}